\documentclass[preprint,prd,nofootinbib,tightenlines,amsmath]{revtex4}
\usepackage{enumerate}
\usepackage{multirow}

\usepackage{epsfig}
\usepackage{graphics,color}      % usual driver
\usepackage{verbatim}
\usepackage{amsmath}   
\catcode`\@=11
\def\sla@#1#2#3#4#5{{%
 \setbox\z@\hbox{$\m@th#4#5$}%
 \setbox\tw@\hbox{$\m@th#4#1$}%
 \dimen4\wd\ifdim\wd\z@<\wd\tw@\tw@\else\z@\fi
 \dimen@\ht\tw@
 \advance\dimen@-\dp\tw@ \advance\dimen@-\ht\z@
 \advance\dimen@\dp\z@
 \divide\dimen@\tw@ \advance\dimen@-#3\ht\tw@
 \advance\dimen@-#3\dp\tw@ \dimen@ii#2\wd\z@
 \raise-\dimen@\hbox to\dimen4{%
 \hss\kern\dimen@ii\box\tw@\kern-\dimen@ii\hss}%
 \llap{\hbox to\dimen4{\hss\box\z@\hss}}}}

\def\slashed#1{%
 \expandafter\ifx\csname sla@\string#1\endcsname{\rm ~Re}lax
{\mathpalette{\sla@/00}{#1}}
\fi}
\def\declareslashed#1#2#3#4#5{%
 \expandafter\def\csname sla@\string#5\endcsname{%
#1{\mathpalette{\sla@{#2}{#3}{#4}}{#5}}}}
 \catcode`\@=12
\declareslashed{}{/}{.08}{0}{D}
 \declareslashed{}{/}{.1}{0}{A}
 \declareslashed{}{/}{0}{-.05}{k}
 \declareslashed{}{/}{.1}{0}{\partial}
 \declareslashed{}{\not}{-.6}{0}{f}

\def\lsim{\mathrel {\vcenter {\baselineskip 0pt \kern 0pt
    \hbox{$<$} \kern 0pt \hbox{$\sim$} }}}
\def\gsim{\mathrel {\vcenter {\baselineskip 0pt \kern 0pt
    \hbox{$>$} \kern 0pt \hbox{$\sim$} }}}

\begin{document}

\baselineskip=15pt
\preprint{}

\title{Impacts of multi-Higgs on the $\rho$ parameter, decays of a neutral Higgs to $WW$ and $ZZ$, and a charged Higgs to $WZ$}

\author{Jian-Yong Cen$^1$\footnote{Electronic address: cenjy@sxnu.edu.cn}, Jung-Hsin Chen{$^2$}\footnote{Electronic address: lewis02030405@gmail.com}, Xiao-Gang He$^{1, 2,3,4}$\footnote{Electronic address: hexg@phys.ntu.edu.tw}, Jhih-Ying Su{$^{2}$}\footnote{Electronic address: b02202013@ntu.edu.tw }, }
\affiliation{$^1$School of Physics and Information Engineering, Shanxi Normal University, Linfen, Shanxi 041004\\
$^2$Department of Physics, National Taiwan University, Taipei 106\\
$^3$T-D. Lee Institute and  School of Physics and Astronomy, Shanghai Jiao Tong University, 800 Dongchuan Road, Shanghai 200240\\
$^4$National Center for Theoretical Sciences, Hsinchu 300}

\date{\today}

\vskip 1cm
\begin{abstract}

In the standard model (SM), the $\rho$ parameter is equal to 1 and  the ratio $\lambda_{WZ}$ of Higgs $\to ZZ$ and Higgs $\to WW$ is also equal to 1 at the tree level. When going beyond the SM with more than one type of Higgs representations these quantities may be different from the SM predictions which can provide crucial information about new physics. There may also exist a certain charged Higgs $h^+$ decays into a $W^+$ and a $Z$. Imposing a custodial symmetry can force the parameter  $\rho$ to be equal to 1 with certain predictions for $\lambda_{WZ}$ and $h^+ \to W^+Z$.  However, imposing $\rho =1$ without custodial symmetry may have different predictions. We show how  differences arise and how to use experimental data to obtain information about the underlying physics in a model with the SM plus a real and a complex $SU(2)_L$ triplets.  

\end{abstract}

\pacs{PACS numbers: }

\maketitle
    
%%%%%%%

\newpage

\section{Introduction}

The discovery of the Higgs boson is a great success of the standard model (SM). Experimental data indicate that the Higgs discovered with a mass of 125 GeV is consistent with that predicted in the SM with just one Higgs doublet of $SU(2)_L\times U(1)_Y$ gauge group~\cite{discovery,PDG:rho}. There are many extensions of the SM in which there are more than one Higgs doublet"s", for example the two Higgs doublets or the minimal SUSY models,  or even different representations than doublet~\cite{multi-Higgs,GM-model,pei-hong}. Experimental searches have not shown signals of new Higgs bosons~\cite{PDG:rho}. 
However, data at present cannot rule out the possibility of beyond SM Higgs boson with a mass beyond the reach of the current data. There may be even charged Higgs bosons. Whether there are new Higgs bosons, neutral or charged, need more experimental data to decide. If indeed new Higgs bosons exist, there are many implications. There are also constrained by various experimental data. Study of modifications due to additional Higgs bosons on the $\rho$ parameter, and properties of a neutral Higgs boson decays into a $WW$ and a $ZZ$ pair can provide interesting information about different models~\cite{lykken}.

In SM, the $\rho$ parameter is equal to 1 and the ratio $\lambda_{WZ}$ of Higgs $\to ZZ$ and Higgs $\to WW$ is also equal to 1 at the tree level. When going beyond the SM with more than one types of Higgs representations these quantities may be different from the SM predictions. Experimental data has shown that the $\rho$ parameter is very close to 1. This provides a stringent constraint on models with multi-Higgs bosons~\cite{PDG:rho}. 
$\rho =1$ may be accidental or may be come from some symmetries, such as custodial symmetry. Higgs boson decay properties may help to distinguish different models. It has been shown that the ratio $\lambda_{WZ}$ of the neutral Higgs boson decays into  a $WW$ pair and a $ZZ$ provide crucial information since in the SM $\lambda_{WZ}$ is predicted to be 1 at the tree level~\cite{lykken}. With multi-Higgs bosons, $\lambda_{WZ}$ may deviate from 1 significantly. Imposing $\rho =1$ with or without custodial symmetry may have different predictions. In models with multi-Higgs boson models, there may be charged Higgs bosons. Some of the simple extensions, such as two Higgs doublet or minimal SUSY models,  do not have tree level contribution to $h^+\to W^+ Z$, and a non-zero contribution can only be generated at loop levels leading to a small decay rate. To have tree level contribution to $h^+\to W^+ Z$ one needs to have two or more non-trivial $SU(2)_L$ representations. If a charged Higgs boson is discovered in the future, its decay modes can also serve to distinguish different extensions of the SM~\cite{charged-Higgs-decay}.

In this paper we study some implications of additional Higgs bosons on the $\rho$ parameter, the parameter $\lambda_{WZ}$ which is the ratio of decay amplitude for a neutral Higgs boson decays into a $WW$ to a $ZZ$ pair, and also a charged Higgs boson $h^+$ decays into a $W^+$ and a $Z$. We show how  differences for these quantities arise and how to use experimental data to obtain information about the underlying physics in a general model with the SM plus a real and a complex $SU(2)_L$ triplets, and also a model with the same Higgs boson multiplets but with a global custodial symmetry. In the following sections, we provide some details of our findings.

\section{The general model}

The model we will study has one doublet $H$,  one complex triplet $\chi$ and a real triplet $\xi$ transforming under $SU(2)_L\times U(1)_Y$ as $(2, -1/2)$, $(3,1)$ and $(3,0)$, respectively. The component fields are given as the following, 
\begin{eqnarray}
H=\left(
\begin{array}{cc} 
h^{0}\\
h^{-}
\end{array}\right)\;,\;\;\;\;\chi=\left(
\begin{array}{cc}
\chi^{+}/\sqrt{2}&\chi^{++}\\
\chi^{0}&-\chi^{+}/\sqrt{2}
\end{array}\right)\;,\;\;\;\;\xi=\left(
\begin{array}{cc}
\xi^{0}/\sqrt{2}&\xi^{+}\\
\xi^{-}&-\xi^{0}/\sqrt{2}
\end{array}\right)\;.
\end{eqnarray}
Since $\xi$ is a real triplet, $\xi^- = (\xi^+)^*$.

The neutral part of each field can develop vacuum expectation values (VEV) and break the electroweak symmetry but keep the $U(1)_{EM}$ symmetry.  We write the fields and their VEVs $v_i$ as
 \begin{eqnarray}
&&h^{0}=\frac{v_H+h_{H}+iI_{H}}{\sqrt{2}}\;,\;\;\;\;\chi^{0}=\frac{v_{\chi}+h_{\chi}+iI_{\chi}}{\sqrt{2}}\;,\;\;\;\;\xi^{0}=v_{\xi}+h_{\xi}\;.
\end{eqnarray}  
  
The terms in the Lagrangian representing the kinetic energy and Higgs potential invariant under the gauge group $SU(2)_{L}\times U(1)_{Y}$ are given by
\begin{equation}
L=(D_{\mu}H)^{\dagger}D^{\mu}H+\frac{1}{2}(D_{\mu}\xi)^\dagger D^{\mu}\xi+(D_{\mu}\chi)^{\dagger}D^{\mu}\chi - V(H, \chi, \xi)\;, \label{kinetic}
\end{equation}
where
\begin{eqnarray}
&&iD_{\mu}H=i\partial_{\mu}H-\frac{g}{2} W_\mu  H+\frac{g'}{2}B_{\mu}H\;,\nonumber\\
&&iD_{\mu}\chi=i\partial_{\mu}\chi-\frac{g}{2}[W_\mu,\chi]-g'B_{\mu}\chi\;,\nonumber\\
&&iD_{\mu}\xi=i\partial_{\mu}\xi-\frac{g}{2}[W_\mu,\xi].
\end{eqnarray}
Here $W_\mu$ and $B_\mu$ are the $SU(2)_L$ and $U(1)_Y$ gauge fields with
\begin{eqnarray}
W_\mu=\left(
\begin{array}{cl}
W_{\mu}^{3}&\sqrt{2}W_{\mu}^{+}\\
\sqrt{2}W_{\mu}^{-}&-W_{\mu}^{3}
\end{array}\right)\;.
\end{eqnarray}
$g$ and $g'$ are the gauge couplings of $SU(2)_L$ and $U(1)_Y$, respectively.

The most general renormalizable Higgs potential is given by
\begin{eqnarray}
V(H, \chi, \xi)&=&\mu_{H}^{2}H^{\dagger}H+\lambda_{H}(H^{\dagger}H)^{2}+\mu_{\chi}^{2} Tr(\chi^{\dagger}\chi)
+\frac{1}{2}\mu_{\xi}^{2} Tr(\xi\xi)\nonumber\\
&+&\lambda_{\chi}(Tr(\chi^{\dagger}\chi))^{2}+\lambda_{\chi}^{\prime}Tr(\chi^{\dagger}\chi \chi^{\dagger} \chi)+ \frac{1}{4}\lambda_{\xi}(Tr(\xi\xi))^{2}\nonumber\\
&+&\frac{{\kappa_{1}}}{2}(H^{\dagger}H) Tr(\xi\xi)+{\kappa_{2}}(H^{\dagger}H)Tr(\chi^{\dagger}\chi)+{\kappa_{3}}(H^{\dagger}\chi\chi^{\dagger}H)\nonumber\\
&+&\frac{{\kappa_{4}}}{4}Tr(\xi\xi)Tr(\chi^{\dagger}\chi)+{ \kappa_{5}}Tr[\xi \chi^{\dagger}]Tr[\xi \chi] \nonumber\\
&+&\mu_{\xi HH} H^\dagger \xi H + \left \{ \mu_{\chi HH} H^{T}\chi H + \lambda H^{T}\chi \xi H + H.C.\right \} + \mu_{\xi\chi\chi} Tr[\chi^\dagger\xi\chi]\;. \label{potential}
\end{eqnarray}

After the spontaneous symmetry breaking, the gauge bosons get masses from Higgs VEVs in the form $(1/2)m^2_Z Z_{\mu}Z^{\mu}$ and $m^2_W W^+_{\mu} W^{- \mu}$.  One obtains~\cite{rho-general}
\begin{eqnarray}  
\rho\equiv\frac{m_{W}^{2}}{\cos^{2}\theta_{W}m_{Z}^{2}}={v^2_H+ 2v_\chi^{2}+4v_{\xi}^2\over v^2_H+4v_{\chi}^2}\;.
\end{eqnarray}

Replacing $v^2_i$ by $2v_i h_i$ from the mass formulae, one obtains $h_i$ couplings to $WW$ and $ZZ$, $g^{h_i}_Z$, and $g^{h_i}_W$ defined for the interaction Lagrangian 
\begin{eqnarray}
L = g_Z^{h_i} (g^2 /\cos^2\theta_W) h_i Z_\mu Z^\mu  + 2 g^{h_i}_W (g^2) h_iW^{+}_\mu W^{-\mu}\;,
\end{eqnarray}
where $g_Z^{h_i}  = Y^2_iv_i $ and $g_{W}^{h_i} = (1/2) (J_i(J_i+1)-Y^2_i)v_i$. Here $J_i$ and $Y_i$ are the $SU(2)_L$ isospin and $U(1)_Y$ hyper-charge of the i\textcolor{red}{-}th Higgs multiplet.

If $h_H$, $h_\chi$ and $h_\xi$ are mass eigenstates, one obtains the ratio $\lambda^{h_i}_{WZ} = g^{h_i}_W/g^{h_i}_Z$ for the three neutral Higgs bosons to be
\begin{eqnarray}
\lambda^{h_H}_{WZ} = 1\;,\;\;\;\;\lambda^{h_\chi}_{WZ} = {1\over 2}\;, \label{no-mixing}
\end{eqnarray}
and $\lambda^{h_\xi}_{WZ}$ would be infinite since $h_\xi$ does not couple to $Z$ boson.

Measurements of Higgs $h_i$ to ZZ and WW couplings can therefore be used to distinguish different Higgs bosons. The correlated analysis of the $\rho$ and $\lambda_{WZ}$ can provide more information about Higgs boson measured in $h \to ZZ$ and $h \to WW$.
However $h_H$, $h_\chi$ and $h_\xi$ are in general not mass eigenstates which can mix with each other. The mixing will change the $\lambda^{h_i}_{WZ}$ predicted in the above. We study how mixing of $h_H$, $h_\chi$ and $h_\xi$ occur in the general model considered here in the following section.

 \section{Potential and Higgs boson masses}

To obtain Higgs boson mass eigenstates, one needs to analyze Higgs potential around the minimum.
We can solve $\mu_{H}$, $\mu_{\chi}$ and $\mu_{\xi}$ at the minimum of the potential to obtain
\begin{eqnarray}
&&\mu_{H}^{2}=\frac{-1}{2}( \sqrt{2} \mu_{\xi HH} v_\xi+2\sqrt{2}\mu_{\chi HH}v_{\chi}+{ \kappa_{1}}v_{\xi}^{2}+{\kappa_{2}}v_{\chi}^{2}+2\lambda v_{\xi}v_{\chi}+2\lambda_{H}v^{2}_H)\;,\nonumber\\
&&\mu_{\chi}^{2}=\frac{-1}{4v_{\chi}}(2\sqrt{2}\mu_{\chi HH}v^{2}_H -2\sqrt{2} \mu_{\xi\chi\chi}v_\xi v_\chi +2{\kappa_{2}}v^{2}_Hv_{\chi}+{\kappa_{4}}v_{\xi}^{2}v_{\chi}+2\lambda v^{2}_Hv_{\xi}+4 \lambda_{\chi}v_{\chi}^{3}+4 \lambda_{\chi}^\prime v_{\chi}^{3})\;,\nonumber\\
&&
\mu_{\xi}^{2}=\frac{-1}{4v_{\xi}}(\sqrt{2} \mu_{\xi HH} v^2_H - \sqrt{2} \mu_{\xi \chi\chi}  v_\chi^{2}+2{\kappa_{1}}v^{2}_Hv_{\xi}+{\kappa_{4}}v_{\xi}v_{\chi}^{2}+2 \lambda v^{2}_Hv_{\chi}+4\lambda_{\xi} v_{\xi}^{3})\;. \label{minimal}
\end{eqnarray}

Inserting the above minimal conditions into the potential, we obtain the mass matrices for the Higgs fields. We list them in the Appendix A.

For real neutral fields $(h_{H}, h_{\xi}, h_{\chi})$, the mass matrix $M^2_{h}$ in Appendix A will mix"ing" all the 3 Higgs bosons. These fields can be expressed as linear combinations of mass eigenstates $h^m_i$ as 
\begin{eqnarray}
\left ( \begin{array}{c}
h_H\\
h_\xi\\
h_\chi
\end{array}
\right ) =
\left ( \begin{array}{ccc}
\alpha_{11}&\alpha_{12}&\alpha_{13}\\
\alpha_{21}&\alpha_{22}&\alpha_{23}\\
\alpha_{31}&\alpha_{32}&\alpha_{33}
\end{array}
\right ) 
 \left ( \begin{array}{c}
h^m_1\\
h^m_2\\
h^m_3
\end{array}
\right )\;.
\end{eqnarray}
Here the matrix $(\alpha_{ij})$ is an orthogonal matrix which has 3 mixing angles in general. 
This leads to very different couplings for $h_i \to WW, ZZ$ compared with the case of no mixing in eq.\ref{no-mixing}.

%For imaginary neutral fields $( I_{H},I_{\chi} )$, we have the mass matrix $M^2_{I}$ shown in Appendix A.
%After removing the would-be Goldstone boson ``eaten'' by Z, the physical field and its mass are given by
%\begin{eqnarray}
%A^0 = (2 v_\chi I_{H}+ v_HI_{\chi})/\sqrt{v^2+4 v^2_\chi}\;,\;\;
%m^2_{A^0} = {- {(\sqrt{2} \mu_{\chi HH} + \lambda v_\xi)(v^2 + 4 v^2_\chi)\over 2 v_\chi}}\;.
%\end{eqnarray}

For later use of discussing $h^+ \to W^+ Z$, we will summarize the singly charged Higgs boson masses and the situation of the mixing briefly here. The mass matrix $M^2_{h^+}$ for the fields $( h^{+}, \xi^{+},\chi^{+})$ is given in Appendix A. One can remove the would-be Goldstone boson ``eaten'' by $W^+$ boson and separate the rest two physical states which we define as
\begin{eqnarray}
&&H^+_3 = {1\over N_2}(-(4 v^2_\xi + 2 v^2_\chi) h^+ + 2 v_H v_\xi \xi^+ - \sqrt{2}v_H v_\chi \chi^+ )\;,\nonumber\\
&&H_5^+ = {1\over N_3} (\sqrt{2} v_\chi \xi^+ + 2 v_\xi \chi^+)\;,\\ 
&&N_2 ^2 = (4v^2_\xi + 2 v^2_\chi)^2 + 4 v^2_Hv^2_\xi + 2 v^2_Hv_\chi^2\;,\nonumber\\
&&N_3^{2} = 4v_\xi^2 + 2 v_\chi^2\;.\nonumber
\end{eqnarray}

In the bases $( H^+_3, H^+_5)$ the elements in the mass matrix $M^{c2}$ is given by
\begin{eqnarray}
M^{c2} = \left ( \begin{array}{cc}
M^{c2}_{11}&M^{c2}_{12}\\
M^{c2}_{12}&M^{c2}_{22}
\end{array}
\right )\;\textcolor{red}{,} \label{charge-m2} \label{mc2}
\end{eqnarray}
shown in Appendix A.

Note that $H^+_3$ and $H^+_5$ are not mass eigenstates. One needs to further diagonalize the mass matrix  in eq.\ref{mc2} into mass eigenstates $H^{m+}_3$ and $H^{m+}_5$
\begin{eqnarray}
\left ( \begin{array}{c}
H^+_3\\
H^+_5
\end{array}
\right ) = 
\left (
\begin{array}{cc}
\cos \delta&\sin\delta\\
-\sin\delta&\cos\delta
\end{array}
\right )
\left ( \begin{array}{c}
H^{m+}_3\\H^{m+}_5
\end{array}
\right )\;, \label{charged-d}
\end{eqnarray}
where
\begin{eqnarray}
\tan2\delta = {2M^{c2}_{12}\over M^{c2}_{22}-M^{c2}_{11}}\;.
\end{eqnarray}

%The doubly charge Higgs boson  $\chi^{++}$ mass $m^2_{\chi^{++}}$ is given by 
%\begin{eqnarray}
%m^2_{\chi^{++}} =- {\mu_{\chi HH} v^2\over \sqrt{2} v_\chi} +\sqrt{2} \mu_{\xi \chi\chi}v_\xi+ {\kappa_3 v^2_H\over 2}  
%- {\lambda v^2_H v_\xi\over 2 v_\chi} - \lambda_{\chi}^\prime v^2_\chi\;.
%\end{eqnarray}

\section{$\rho = 1$ and $\lambda^{h_i}_{WZ}$}

When discussing about $\lambda^{h_i}_{WZ}$ one should always use mass eigenstates because they are the states being measured.
We have
\begin{eqnarray}
\lambda^{h^m_i}_{WZ} = {v_H \alpha_{1i} + 4 v_\xi \alpha_{2i} + 2 v_\chi \alpha_{3i}\over v_H \alpha_{1i} + 4 v_\chi \alpha_{3i}}
= 1 + {2 (2 v_\xi \alpha_{2i}  - v_\chi \alpha_{3i})\over v_H \alpha_{1i} + 4 v_\chi \alpha_{3i}}\;.
\end{eqnarray}

The condition to obtain $\lambda^{h^m_i}_{WZ} =1$ is that $2 v_\xi \alpha_{2i}  - v_\chi \alpha_{3i}=0$.  
Because $\sum_i \alpha_{2i}\alpha_{3i} =0$, this condition also tells that it is not possible to have all $\lambda^{h^m_i}_{WZ}=1$. But with two of them to be equal to 1 is possible.

Experimental data on precision electroweak measurement constrain $\rho$ to be very close to one with 
$\rho =1.00037\pm 0.00023$\cite{PDG:rho}. It may be interesting to impose $\rho =1$ to see what consequences are.
In the general model $\rho$ deviates from 1. Imposing $\rho = 1$, we have
\begin{eqnarray}
\rho\equiv\frac{m_{W}^{2}}{\cos^{2}\theta_{\textcolor{red}{W}}m_{Z}^{2}}={v^2_H+ 2v_\chi^{2}+4v_{\xi}^2\over v^2_H+4v_{\chi}^2} = 1\;,
\end{eqnarray}
which forces the relation
\begin{eqnarray}
v_\xi = {v_\chi \over \sqrt{2}}\;. \label{aligan}
\end{eqnarray}
With this relation, the VEV of $\chi$ and $\xi$ need not to be very small which may have some interesting phenomenology, such as Type II seesaw models.  In this case
\begin{eqnarray}
\lambda^{h^m_i}_{WZ} = 1 +{ 2 v_\chi(  \sqrt{2} \alpha_{2i} - \alpha_{3i})\over v_H \alpha_{1i} + 4 v_\chi \alpha_{3i}}\;.
\end{eqnarray}
Again it is not possible to have all $\lambda^{h^m_i}_{WZ}$ to be equal to 1.

From $M^2_h$ in the Appendix A, one can see that even with the condition $v_\xi =v_\chi/\sqrt{2}$, there are still a large parameter space where $\lambda^{h^m_i}_{WZ}$ are not completely fixed.  For example, with $\alpha_{13} = 0$ which can be realized in the general model, one can write
\begin{eqnarray}
\left ( \begin{array}{c}
h_H\\
h_\xi\\
h_\chi
\end{array}
\right ) =
\left ( \begin{array}{ccc}
\cos \alpha&\sin \alpha&0\\
-\cos\gamma \sin \alpha&\cos\gamma\cos\alpha&\sin\gamma\\
\sin\gamma \sin \alpha&-\sin\gamma\cos \alpha&\cos\gamma
\end{array}
\right ) 
 \left ( \begin{array}{c}
h^m_1\\
h^m_2\\
h^m_3
\end{array}
\right )\;.
\end{eqnarray}

One obtains
\begin{eqnarray}
&&\lambda_{WZ}^{h_1^m}=1- {2v_\chi(\sqrt{2}\cos\gamma+\sin\gamma) \over v_H \cot\alpha+4v_\chi \sin\gamma}\;,\nonumber\\
&&\lambda_{WZ}^{h_2^m}=1+{2v_{\chi}(\sqrt{2}\cos\gamma+\sin\gamma)\over v_H \tan\alpha - 4v_{\chi}\sin\gamma}\;,\nonumber\\
&&\lambda^{h^m_3}_{WZ} = {1\over 2} + {1\over \sqrt{2}} \tan \gamma\;. \label{parameterization}
\end{eqnarray}
There are 3 free parameters which lead to wide ranges for $\lambda^{h^m_i}_{WZ}$. Experimental measurements can help to narrow down the allowed ranges and determine model parameters.

The above scenario can be achieved by requiring the following
\begin{eqnarray}
&&\tan\gamma=-\frac{2 \mu_{\chi HH}+\sqrt{2} \kappa_{2}v_{\chi}+\lambda v_{\chi}}{\mu_{\xi HH}+\kappa_{1}v_{\chi} + \sqrt{2}\lambda v_{\chi}}\;,\\
&&{1- \tan^2\gamma\over \sqrt{2} \tan\gamma} = { (\sqrt{2}\mu_{\chi HH}-\mu_{\xi HH})v^2_H + \mu_{\xi \chi\chi}v^2_\chi  
+2(\lambda_\xi  -2 \lambda_\chi -2 \lambda_\chi^\prime)v^3_\chi -\lambda v^2_H v_\chi /\sqrt{2}
\over v_\chi (2\mu_{\xi \chi\chi}v_\chi  -\kappa_4 v^2_\chi -\sqrt{2}\lambda v_H^2)}\;.\nonumber
\end{eqnarray}
For given $v_i$, by varying the parameters in the potential, $\tan\gamma$ can have consistent solution and can take a wide range of values.

\section{Comparison with Georgi-Machacek model}

In the Georgi-Machacek (GM) model, there are same number of Higgs fields as the general model discussed in the previous sections. However a custodial global $SU(2)_L\times SU(2)_R$ symmetry is imposed on the Higgs potential~\cite{GM-model}. The Higgs fields $H$ is written in a form $\Phi$, and $\xi$ and $\chi$ are grouped into one multiplet $\Delta$, which transform under the custodial symmetry as (2, 2) and (3, 3) multiplets, respectively. They are given as follows
 \begin{eqnarray}
 \Phi=\left (\begin{array}{cc} h^{0*}&h^+\\-h^{+*}&h^0
 \end{array}
 \right )\;,\;\;\Delta =\left (\begin{array}{ccc}\chi^{0*}&\xi^+ &\chi^{++}\\-\chi^{+*}&\xi^0&\chi^+\\
 \chi^{++*}&-\xi^{+*}&\chi^0\end{array}
 \right )\;.
 \end{eqnarray}
 
The Higgs potential respecting the custodial symmetry is given by~\cite{GM-model,chiang}
\begin{eqnarray}
V_{GM} &=& {1\over 2} m^2_1 Tr(\Phi^\dagger \Phi) + {1\over 2} m^2_2 Tr(\Delta^\dagger \Delta) + \lambda_1(Tr(\Phi^\dagger \Phi))^2 + \lambda_2 (Tr(\Delta^\dagger \Delta))^2\nonumber\\
&+&\lambda_3 Tr(\Delta^\dagger \Delta \Delta^{\textcolor{red}{\dagger}}\Delta)+\lambda_4 Tr(\Phi^\dagger \Phi)Tr(\Delta^\dagger \Delta) + \lambda_5 Tr(\Phi^\dagger {\sigma^a\over 2}\Phi {\sigma^b\over 2}) Tr(\Delta^\dagger T^a\Delta T^b)\nonumber\\
&+& \mu_1 Tr(\Phi^\dagger {\sigma^a\over 2} \Phi {\sigma^b\over 2} )Tr(P^\dagger \Delta P) + \mu_2 Tr(\Delta T^a \Delta T^b)Tr(P^\dagger \Delta P)\;,
\end{eqnarray}
where $\sigma^a$ and $T^a$ are the generators $SU(2)$ in the 2- and 3 dimensions, and 
\begin{eqnarray}
P = {1\over \sqrt{2}} \left ( \begin{array}{ccc}
-1& i& 0\\0&0&\sqrt{2}\\1&i&0\end{array}
\right )\;.
\end{eqnarray}

The minimal condition for above potential with a non-zero VEV, $v_H$,  for $H$ is
\begin{eqnarray}
m^2_1 = -{1\over v_H} (4 \lambda_1 v^3_H  + 2 \lambda_4 v_H(v^2_\xi + v^2_\chi) + {\lambda_5\over 4} (4\sqrt{2} v_H v_\xi v_\chi + 2 v_H v^2_\chi) +{\mu_1\over 4}(2 v_H v_\xi + 2 \sqrt{2}v_H v_\chi))\;,\nonumber\\
\end{eqnarray}
and the minimal conditions for non-zero VEVs $\xi$ and $\chi$ if they 
are different, $m^2_2$ needs to satisfy both
\begin{eqnarray}
m^2_2 = -{1\over v_\xi} (4\lambda_2 v_\xi(v^2_\xi +v^2_\chi) + 4 \lambda_3 v^3_\xi + 2 \lambda_4 v^2_H v_\xi +\lambda_5\frac{v^2_H v_\chi} {\sqrt{2}}  + {\mu_1 v^2_H \over 4} + 3 \mu_2 v^2_\chi)\;,\nonumber
\end{eqnarray}
and
\begin{eqnarray}
m^2_2 = -{1\over v_\chi} (4\lambda_2 v_\chi(v^2_\xi +v^2_\chi) + 2\lambda_3 v^3_\chi+ 2 \lambda_4 v^2_H v_\chi +{\lambda_5 v^2_H  (2\sqrt{2}v_\xi + 2 v_\chi) + \sqrt{2}\mu_1 v^2_H \over 4} + 6 \mu_2 v_\xi v_\chi)\;.\nonumber
\end{eqnarray}
Note that for $v_\xi \neq v_\chi/\sqrt{2}$, there is no consistent solution. The custodial symmetry forces $v_\xi = v_\chi/\sqrt{2}$ to have a consistent solution. This guarantees $\rho = 1$ at the tree level~\cite{GM-model}.  

In this model, $h^m_1 = h$, 
$h^m_2 = H_1^0$ and $h^m_3 = H^0_5$  are linear combinations of $h_H$, $h_\xi$ and $h_\chi$ given by
\begin{eqnarray}
&&\left ( \begin{array}{c}
h_H\\
h_\xi\\
h_\chi
\end{array}
\right ) =
\left ( \begin{array}{ccc}
\cos \alpha&\sin \alpha&0\\
-\sqrt{{1\over 3}}\sin \alpha&\sqrt{{1\over 3}}\cos\alpha&-\sqrt{{2\over 3}}\\
-\sqrt{{2\over 3}} \sin \alpha&\sqrt{{2\over 3}}\cos \alpha&\sqrt{{1\over 3}}
\end{array}
\right ) 
 \left ( \begin{array}{c}
h^m_1\\
h^m_2\\
h^m_3
\end{array}
\right )\;.
\end{eqnarray}
The angle $\alpha$ is given by
\begin{eqnarray}
\tan 2\alpha = {2m^2_{12}\over m^2_{22} -m^2_{11}}\;,
\end{eqnarray}
%\begin{eqnarray}
%\left (\begin{array}{cc}
%m^2_{h^m_1}&0\\0&m^2_{h^m_2}
%\end{array}
%\right )
%= \left (\begin{array}{cc}
%\cos\alpha&-\sin\alpha\\\sin\alpha&\cos\alpha
%\end{array}
%\right )
%\left (\begin{array}{cc}
%m^2_{11}&m^2_{12}\\m^2_{12}&m^2_{22}
%\end{array}
%\right )
% \left (\begin{array}{cc}
%\cos\alpha&\sin\alpha\\-\sin\alpha&\cos\alpha
%\end{array}
%\right )\;,
%\end{eqnarray}
with
\begin{eqnarray}
&&m^2_{11}= 8 \lambda_1 v^2_H\;,\;\;
m^2_{12}=m^2_{21}={\sqrt{3} v_H\over 2}(2\sqrt{2}(2\lambda_4 + \lambda_5)v_\chi +\mu_1)\;,\nonumber\\
&&m^2_{22}= 4(3\lambda_2 + \lambda_3) v_\chi^2 -{\mu_1 v^2_H\over 2\sqrt{2} v_\chi} + 3\sqrt{2}\mu_2 v_\chi\;.
\end{eqnarray}
The mass of $h^m_3 = H_5$ is given by
\begin{eqnarray}
m^2_{H_5} = 4\lambda_3 v_\chi^2 - {3\lambda_5 v^2_H\over 2} -{\mu_1 v^2_H\over 2\sqrt{2} v_\chi} - 6\sqrt{2}\mu_2 v_\chi\;.
\end{eqnarray}

The two physical singly charged Higgs bosons $H_3^+$ and $H_5^+$ and their masses are 
\begin{eqnarray}
&&h^{m+}_2 = H_3^+ = {2\sqrt{2} v_\chi h^+ - v_H \xi^+ - v_H \chi^+ \over \sqrt{2 v_{H}^2 + 8 v^2_\chi}}\;,\;\;m^2_{H_3^+} = -{(2\lambda_5 v_\chi + \sqrt{2} \mu_1)(v^2_H + 4 v^2_\chi)\over 4 v_\chi}\;,\\
&&h^{m+}_3 = H_5^+ = {1\over \sqrt{2}}(\xi^+-\chi^+)\;,\;\;m^2_{H_5^+}=m^2_{\chi^{++}}= 4\lambda_3 v_\chi^2 - {3\lambda_5 v^2_H\over 2} -{\mu_1 v^2_H\over 2\sqrt{2} v_\chi} - 6\sqrt{2}\mu_2 v_\chi\;.\nonumber
\end{eqnarray}

The doubly charged Higgs boson $\chi^{++} = H_5^{++}$ has a mass equal to $m^2_{H_5}$.

One would obtain~\cite{lykken,chiang}
\begin{eqnarray}
&&\lambda^{h^m_1}_{WZ} = 1\;,\;\;\;\;\lambda^{h^m_2}_{WZ} = 1\;,\;\;\;\;\lambda^{h^m_3}_{WZ} = -{1\over 2}\;.
\end{eqnarray}

The general model, discussed in previous sections, is very different from the GM model. For example even with
$v_\xi = v_\chi/\sqrt{2}$, the minimal conditions do not lead to the same as GM model structure with
\begin{eqnarray}
&&\mu^2_H = -\lambda_{H} v^2_H -{1\over 4}(\kappa_1 +2\kappa_2)v^2_\chi - {1\over \sqrt{2}} \lambda v^2_\chi -{1\over 2} (\mu_{\xi HH} +2\sqrt{2} \mu_{\chi HH}) v_\chi\;,\nonumber\\
&&\mu^2_\xi = -{1\over 2} \lambda_\xi  v^2_\chi- {1\over 4} (2\kappa_1v^2_H +\kappa_4v^2_\chi) - { 1\over \sqrt{2}}\lambda v^2_H
-{1\over 2 v_\chi}(\mu_{\xi HH} v^2_H - \mu_{\xi \chi\chi} v^2_\chi)\;,\\
&&\mu^2_\chi = - (\lambda_\chi + \lambda_\chi^{\prime})v^2_\chi - {1\over 8} (4 \kappa_2 v^2_H + \kappa_4 v^2_\chi) - {1\over 2 \sqrt{2}} \lambda v^2_H -{1\over  2 v_\chi}(\sqrt{2} \mu_{\chi HH} v^2_H - \mu_{\xi \chi\chi} v^2_\chi)\;.\nonumber
\end{eqnarray}
In the GM model, it would imply $\mu^2_\xi = \mu^2_\chi$. But this is not always true for the general model.

Additional constraints need to be applied to reduce the general model to the GM model. We find that by
setting $v_\xi = v_\chi/\sqrt{2}$, and
\begin{eqnarray}
&&\mu_H^2 = m^2_1\;,\;\;\mu^2_\xi = m^2_2\;,\;\;\mu^2_\chi = m^2_2\;,\nonumber\\
&&\lambda_H = 4 \lambda_1\;,\;\;\lambda_\xi = 4\lambda_2+ 4\lambda_3\;,\;\;\lambda_\chi = 4 \lambda_2 + 6\lambda_3\;,\lambda^{\prime}_{\chi}=-4\lambda_{3}\nonumber\\
&&\kappa_1 = 4 \lambda_4\;,\;\;\kappa_2 = 4\lambda_4 + \lambda_5\;,\;\;\kappa_3 = -2\lambda_5\;,\;\;\kappa_4 = 16\lambda_2\;,\;\;\kappa_5 = 4 \lambda_3\;\nonumber\\
&&\mu_{\xi HH} = {\mu_1\over \sqrt{2}}\;,\;\;\mu_{\chi HH} ={\mu_1\over 2}\;,\;\;\mu_{\xi \chi\chi} = - 6\sqrt{2} \mu_2\;,\lambda=\sqrt{2}\lambda_{5}\;, \label{conditions}
\end{eqnarray}
the model leads to the same mass matrices for the Higgs bosons in the potential given by eq.\ref{potential} and those in GM model~\cite{yageu}. 
The reduced potential in given in Appendix B. Note that  conventions in the general model and the GM model are different. 
To obtain the same results in forms, one needs to replacing $I_\chi \to -I_\chi$, $h^+_\chi \to - h^+_\chi$ in the bases for eqs. \ref{imaginary} and \ref{w-h} in Appendix A.

In the general model even with $v_\xi = v_\chi/\sqrt{2}$,  $\lambda^{h^m_i}_{WZ}$ can take a much wider ranges than those predicted in the GM model. This can be easily shown to be true using eq.\ref{parameterization}.
Only with $\tan \gamma = -\sqrt{2}$ and $\cos\gamma = 1/\sqrt{3}$,  the model predicts the same $\lambda^{h^m_i}_{WZ}$ as those in the GM model. 

Experimentally only one Higgs boson $h^{exp}$ with a mass 125 GeV has been discovered. The value $\lambda^{h^{exp}}_{WZ}$ is consistent with SM prediction. It is also consistent with GM model prediction if $h$ or $H_3$ is $h^{exp}$. If in the future more Higgs bosons will be discovered with different values for  $\lambda^{h^m_i}_{WZ}$, this will provide crucial information about the underlying model. The general model discussed here provides a concrete example for $\lambda^{h^m_i}_{WZ}$ to be deviate from 1 and also deviate from GM predictions.

\section{$h^+_i \to W^+ Z$ with Multiplet Higgs fields}

The process for a charged Higgs boson $h^+$ decays into a $W^+$ boson and  $Z$ boson, $h^+ \to W^+ Z$, can only occur beyond the SM since there is no charged Higgs at all. When going beyond the SM, such as two Higgs doublet model or minimal SUSY model, there is a physical charged Higgs. However\textcolor{red}{,} in two Higgs doublet or minimal SUSY models there are no tree level contribution to $h^+\to W^+ Z$, and a non-zero contribution can only be generated at loop levels leading to a small decay rate. To have tree level contribution to $h^+\to W^+ Z$ one needs to have two or more non-trivial $SU(2)_L$ representations, such as the model we are considering.  At present now charged Higgs boson has been detected\cite{PDG:rho}. Should in the future $h^+\to W^+ Z$ be discovered, this process can also serve to distinguish different multi-Higgs models beyond SM~\cite{rho-general, charged-Higgs-decay, chiang}. We now provide some details for the general model and the GM model.

Expanding the kinetic energy terms in eq.\ref{kinetic}, one can find the would-be Goldstone model ``eaten'' by $W^+$ and the physical charged Higgs degrees of freedom couplings to $W^+ Z$. We have
\begin{eqnarray}
L = {v_Hg^2\over 2 c_W}(1-c^2_W) h^+ W^-_\mu  Z^\mu - {\sqrt{2} v_\chi g^2\over 2 c_W}(2-c^2_W) h^+_\chi W^-_\mu Z^\mu - c_W v_\xi g^2h^+_\xi W^-_\mu Z^\mu\;.
\end{eqnarray}

Removing the would-be Goldstone mode couplings, we obtain the physical charged degrees of freedom $H^+_3$ and $H^+_5$ couplings to $W^+Z$ to be
\begin{eqnarray}
L &=& {g^2\over 2 c_W} {v_H(2v^2_\chi - 4 v^2_\xi)\over N_2} H^+_3 W^-_\mu Z^\mu - {g^2\over 2c_W} {4\sqrt{2}  v_\chi v_\xi\over {N_3}} H^+_5 W^-_\mu Z^\mu\;,
\end{eqnarray}
in the GM model. Since in this model $v_\xi = v_\chi/\sqrt{2}$, only $H^+_5$ can decay into $W^+Z$.

In the general model, there is a mixing between $H^+_3$ and $H^+_5$, one would have
\begin{eqnarray}
L &=& ({g^2\over 2 c_W} {v_H(2v^2_\chi - 4 v^2_\xi)\over N_2} \cos\delta  + {g^2\over 2c_W} {4\sqrt{2}  v_\chi v_\xi\over {N_3}} \sin\delta )H_3^{m+} W^-_\mu Z^\mu\nonumber\\
&+&({g^2\over 2 c_W} {v_H(2v^2_\chi - 4 v^2_\xi)\over N_2} \sin\delta  - {g^2\over 2c_W} {4\sqrt{2}  v_\chi v_\xi\over {N_3}} \cos\delta )H_5^{m+} W^-_\mu Z^\mu\;.
\end{eqnarray}
Note that with $v_\xi = v_\chi/\sqrt{2}$ constraint, one would find the coupling for the $H_3$ term vanishes. But since in general $sin\delta$ is not zero, both $H^{m+}_{3}$ and $H^{m+}_5$ can decay into $WZ$.

In the GM model one has
\begin{eqnarray}
&&\Gamma(H^+_5 \to W^+ Z) ={g^4 v_{\chi}^4 F(m_W, m_Z, m_{H_5})\over 2 \pi c_W^2 N_{3}^{2}m_{H_{5}}}\;,\\
&&F(m_1, m_2, m_3) = \sqrt{(1-{(m_1+m_2)^{2}\over m_3^2})(1-{(m_1-m_2)^2\over m_3^2})} (1+{(m_1^2+m_2^2-m_3^2)^{2}\over 8m_1^2m_2^2})\;.\nonumber
\end{eqnarray}

In the more general model, $H^+_3$ and $H^+_5$ are not mass eigenstates, one needs to further diagonalized them as given in eq.\ref{charged-d}, we have
\begin{eqnarray}
&&\Gamma(H^{m+}_3 \to W^+ Z) ={g^4v_\chi^4\sin^2\delta F(m_W, m_Z, m_{H^{m+}_3})\over 2\pi c_W^2N_3^2 m_{H^{m+}_3}}(1+\Delta C \cot\delta)^2\;,\nonumber\\
&&\Gamma(H^{m+}_5 \to W^+ Z) ={g^4v_{\chi}^4\cos^2\delta F(m_W, m_Z, m_{H^{m+}_5})\over 2\pi c_W^2N_3^2 m_{H^{m+}_5}}(1-\Delta C \tan\delta)^2\;,
\end{eqnarray}
where $\Delta C = (N_3/N_2)(v_H(2v^2_\chi - 4 v^2_\xi)/4\sqrt{2}v_\xi v_\chi)$ which represents the contribution when $v_\xi$ deviates from $v_\chi/\sqrt{2}$ in the general model.

The above can be used to distinguish different models, that is if the model satisfy the custodial symmetry, there is only one Higgs $H^+_5$ can decay into $W^+Z$, but the more general model, even with $\rho =1$, can have two singly charged Higgs bosons decay into $W^+Z$ the ratio of couplings determines the mixing $\tan \delta$ between $H^+_3$ and $H^+_5$. In the more general model, the ratio depends also the ratios of the VEVs.

\section{Conclusions}

In this paper we have studied the impacts of multi-Higgs on the $\rho$ parameter, the ratio $\lambda_{WZ}$ of the decay width for a neutral Higgs to a $WW$ pair and to   a $ZZ$ pair, and a charged Higgs decays to $W^+Z$. We have performed a detailed analysis for a general model with the SM plus a real and a complex $SU(2)_L$ triplets and also the Georgi-Machacek model which have the same additional Higgs multiplets but with a custodial symmetry. If the complex triplet VEV $v_\chi$ and the real triplet VEV $v_\xi$ do not satisfy the relation $v_\xi = v_\chi/\sqrt{2}$ enforced by the custodial symmetry in the GM model, $\rho$ is not equal to one. Imposing this relation in the 
general model, one can force $\rho$ to be 1, but the predicted $\lambda_{WZ}$ for physical neutral Higgs can still be totally different than those predicted in the GM model.  In these models, there are charged Higgs bosons. Some of them can have tree level $h^+\to W^+ Z$ decays. The general model again have very different predictions than those of the GM model.
Precise measurements of $\lambda_{WZ}$ and $h^+ \to W^+ Z$ can provide crucial information for different models if indeed the additional  Higgs bosons exist.
 
\begin{acknowledgments}

\end{acknowledgments}

This work was supported in part by the MOST (Grant No. MOST104-2112-M-002-015-MY3 and 106-2112-M-002-003-MY3 ), and in part by Key Laboratory for Particle Physics,
Astrophysics and Cosmology, Ministry of Education, and Shanghai Key Laboratory for Particle
Physics and Cosmology (Grant No. 15DZ2272100), and in part by the NSFC (Grant Nos. 11575111 and 11735010).

\appendix

\section{Higgs boson mass matrices in the general model}

Inserting the minimal conditions in eq.\ref{minimal} into eq.\ref{potential}, one obtains the mass matrices for the Higgs fields. We list them below.

For real neutral fields $( h_{H}, h_{\xi}, h_{\chi})$ whose mass matrix $M^2_{h}$ have the following elements 
\begin{eqnarray}
&&m^2_{11}= 2\lambda_{H}v^{2}_H\;,\nonumber\\
&&m^2_{12} = m^2_{21} = v_H(\frac{\mu_{\xi HH}}{\sqrt{2}}+\lambda v_{\chi}+{\kappa_{1}}v_{\xi})\;,\nonumber\\
&&m^2_{13} = m^2_{31}= v_H({ {\sqrt{2}\mu_{\chi HH}}}+{{\kappa_{2}}v_{\chi}}+{\lambda v_{\xi}})\;,\nonumber\\
&&m^2_{22} = \frac{- \sqrt{2}\mu_{\xi HH}  v^{2}_H +\sqrt{2}\mu_{\xi \chi\chi} v^{2}_\chi +8 \lambda_{\xi}v_{\xi}^3 - 2 \lambda v_\chi v^2_H}{4v_{\xi}}\;,\nonumber\\
&&m^2_{23} =m^2_{32} = {1\over 2} (\lambda v^{2}_H -{\sqrt{2}}\mu_{\xi \chi \chi} v_\chi +{\kappa_{4}}v_\xi v_{\chi})\;,\nonumber\\
&&m^2_{33} = {\frac{-\sqrt{2}\mu_{\chi HH}v^{2}_H-\lambda v^{2}_Hv_{\xi}+4(\lambda_{\chi}+\lambda^{\prime}_{\chi})v_{\chi}^{3}}{2v_{\chi}}}\;.
\label{real-mass}
\end{eqnarray}

For imaginary neutral fields $( I_{H},I_{\chi} )$, we have the mass matrix $M^2_{I}$\\
\begin{eqnarray}
%M^2_{I}=
\left(
\begin{array}{cc}
-2v_{\chi}(\sqrt{2}\mu_{\chi HH}+\lambda v_\xi)&- v_H(\sqrt{2}\mu_{\chi HH} +\lambda {v_\xi})\\
-v_H(\sqrt{2}\mu_{\chi HH} +\lambda {v_\xi})&-\frac{v^{2}_H(\sqrt{2}\mu_{\chi HH}+\lambda v_\xi)}{2 v_{\chi}}
\end{array}\right)\;. \label{imaginary}
\end{eqnarray}

The above mass matrix has a zero determinant implying a massless eigenstate which is $G_Z = (v_HI_{H}-2v_{\chi}I_{\chi})/\sqrt{v^2_H+4 v^2_\chi}$.  The field $A^0 = (2 v_\chi I_{H}+ v_HI_{\chi})/\sqrt{v^2_H+4 v^2_\chi}$ is a physical scalar with a mass given by
\begin{eqnarray}
m^2_{A^0} = {- {(\sqrt{2} \mu_{\chi HH} + \lambda v_\xi)(v^2_H + 4 v^2_\chi)\over 2 v_\chi}}\;.
\end{eqnarray}

For the singly charged fields $( h^{+}, \xi^{+},\chi^{+})$, we have the elements in the $3\times 3$ singly charged mass matrix $M^2_{h^+}$
\begin{eqnarray}
&&m^2_{11}=\frac{-2\sqrt{2}\mu_{\xi HH} v_\xi -{v_{\chi}(2\sqrt{2}\mu_{\chi HH} -2\lambda v_{\xi}+{\kappa_{3}}v_{\chi})}}{2}\;,\nonumber\\
&&m^2_{22} = {\frac{-\sqrt{2}\mu_{\xi HH}v^{2}_H-v_{\chi}(2\lambda v_{H}^{2}+2{\kappa_{5}}v_{\xi}v_{\chi}+\sqrt{2}
\mu_{\xi \chi \chi}v_{\chi}) }{4v_{\xi}}}\;,\nonumber\\
&&m^2_{33} = {\frac{-2v^{2}_H(\sqrt{2}\mu_{\chi HH}+\lambda v_{\xi})+v_{\chi}({\kappa_{3}}v^{2}_H+2\sqrt{2}\mu_{\xi \chi \chi}v_{\xi}+4{\kappa_{5}}v_{\xi}^{2})}{4v_{\chi}}}\;.\nonumber\\
&&m^2_{12} = m^2_{21} ={\frac{v_H(\sqrt{2}\mu_{\xi HH}+\lambda v_{\chi})}{2}}\;,\nonumber\\
&&m^2_{13} = m^2_{31} = {\frac{v_H(-4\mu_{\chi HH}  +\sqrt{2}{\kappa_{3}} v_{\chi})}{4}}\;,\nonumber\\
&&m^2_{23} = m^2_{32} = {\frac{-\sqrt{2}\lambda v^{2}_H+2v_{\chi}(\mu_{\xi \chi \chi}+\sqrt{2}{\kappa_{5}}v_{\xi})}{4}}\;.
\label{w-h}
\end{eqnarray}

We have  a massless eigenstate $(v_Hh^{+} + 2 v_{\xi}\xi^{+}-\sqrt{2}v_{\chi}\chi^{+})/\sqrt{v^2_H + 4 v^2_\xi + 2 v^2_\chi}$ which is the would-be Goldstone boson ``eaten'' by $W^+$ and its hermitian conjugate field is the one ``eaten'' by $W^-$. 
After removing the Goldstone model in the above equation, the mass matrix in the bases $( H^+_3, H^+_5)$ is given by 
\begin{eqnarray}
M^{c2}=\left ( \begin{array}{cc}
M^{c2}_{11}&M^{c2}_{12}\\
M^{c2}_{12}&M^{c2}_{22}
\end{array}
\right )\;. \label{charge-m2}
\end{eqnarray}
with
\begin{eqnarray}
&&M^{c2}_{11} = -{(v^2_H+4v_\xi^2 + 2 v_\chi^2)(2\sqrt{2}\mu_{\xi HH}v_\xi + v_\chi (2\sqrt{2} \mu_{\chi HH} + 2 \lambda v_\xi - \kappa_3 v_\chi)\textcolor{red}{)} \over 4 (2v^2_\xi +  v^2_\chi)}\;,\nonumber\\
&&M^{c2}_{12} = -{v_H\sqrt{v_{H}^2 + 4v^2_\xi + 2 v_\chi^2}(-4 \mu_{\chi HH} v_\xi + v_\chi (2\mu_{\xi HH} + \sqrt{2}(\kappa_3 v_\xi + \lambda v_\chi)))\over 4(2 v^2_\xi + v^2_\chi)}\;,\nonumber\\
&&M^{c2}_{22} = -{1\over 4 v_\xi v_\chi (2v^2_\xi +  v^2_\chi)}(v^2_H(4\sqrt{2} \mu_{\chi HH} v_\xi^3 - 2 \kappa_3 v_\xi^3 v_\chi + \sqrt{2} \mu_{\xi HH} v^3_\chi\nonumber\\
&&\;\;\;\;\;\;\;\;\;\;\;+ 2 \lambda(2 v_\xi^4 + 2 v^2_\xi v^2_\chi + v^4_\chi) ) -\sqrt{2} \mu_{\xi \chi\chi} v_\chi (2v^2_\xi + v^2_\chi)^2 \nonumber\\
&&\;\;\;\;\;\;\;\;\;\;\;-v_\chi\kappa_5 (8v^5_\xi +8v^3_\xi v^2_\chi+2 v_\xi v^4_\chi )) \;.
\end{eqnarray}

The doubly charge Higgs boson  $\chi^{++}$ mass $m^2_{\chi^{++}}$ is given by 
\begin{eqnarray}
m^2_{\chi^{++}} =- {\mu_{\chi HH} v^2_H\over \sqrt{2} v_\chi} +\sqrt{2} \mu_{\xi \chi\chi}v_\xi+ {\kappa_3 v_{H}^2\over 2}  - {\lambda v^2_H v_\xi\over 2 v_\chi} - \lambda_{\chi}^\prime v^2_\chi\;.
\end{eqnarray}

\section{Reduction of Higgs potential from general to GM models}

With conditions given in eq.\ref{conditions}, the Higgs potential in eq.\ref{potential} reduces to the potential in GM model which can be written as 
\begin{eqnarray}
V=V_{GM} &=& m^2_1 H^\dagger H + {1\over 2} m^2_2 Tr(\xi \xi) + m^2_2 Tr(\chi^\dagger \chi)\;,\nonumber\\
&+&4\lambda_1(H^\dagger H)^2 + \lambda_2 \left(Tr(\xi\xi) + 2 Tr (\chi^\dagger \chi)\right )^2\nonumber\\
&+&\lambda_3\left ((Tr(\xi\xi))^2 + 4 Tr(\xi \chi^\dagger)Tr(\xi \chi) + 6 (Tr(\chi^\dagger \chi))^2 - 4 Tr(\chi^\dagger \chi \chi^\dagger \chi)\right )\nonumber\\
&+&\lambda_4\left (2 (H^\dagger H)(Tr(\xi\xi) + 2 Tr(\chi^\dagger \chi) )\right)\nonumber\\
& +& \lambda_5\left ( (H^\dagger H) Tr(\chi^\dagger \chi) - 2H^\dagger \chi \chi^\dagger H + (\sqrt{2} H^\dagger \chi \xi H + H.C.) \right )\nonumber\\
&+& \mu_1 \left ({1\over \sqrt{2}}H^\dagger \xi H + ({1\over 2} H^T \chi H + H.C.)\right ) + \mu_2 \left (-6\sqrt{2} Tr(\chi^\dagger \xi \chi)\right )\;.
\end{eqnarray}

\end{document}